# PI Controller for Active Twin-Accumulator Suspension with Optimized Parameters Based on a Quarter Model

Mohamed A. Hassan [*][1], Ali M. Abd-El-Tawwab [1], k. A. Abd El-gwwad [1] and M. M. M. Salem [1]

[1] *Automotive and Tractors Engineering Department, Faculty of Engineering, Minia University, El-Minia 61111, Egypt.*

[*]*Corresponding Authors: E-mail: eng.mohamed_hassan@mu.edu.eg*

**Abstract:**

This paper is primarily studying the behaver of the twin-accumulator suspension over the conventional passive system focusing on ride quality behavior and road holding. Therefore, a dynamic modeling of passive and twin-accumulator suspension for a quarter model is constructed. MATLAB Simulink environment is used to develop the suspension models. The simulation is applied with two different road disturbances, namely, step input and random input to disturb the suspension system. The optimum solution is obtained numerically by utilizing a multi-objective evolutionary strategy algorithm and employing a cost function that seeks to minimize the RMS value of the body acceleration, the suspension displacement as well as the dynamic tire load. Moreover, in this work, an active suspension system with PI controller is presented in order to improve the suspension performance criteria. The simulation results of passive, optimized twin accumulator suspension and active suspension consist of body displacement, wheel deflection, vehicle body acceleration, suspension travel and dynamic tire load are compared and analyzed. Results show that the twin-accumulator suspension system gives worthwhile improvements in ride behavior compared with the passive suspension. Finally, it can be observed that the performance of body displacement and wheel displacement can be improved by using the proposed PI controller.

## 1. Introduction:

The main target of an automotive suspension system is to provide vehicle support, stability and directional control during handling maneuvers and provide effective isolation from road disturbance. Therefore, suspension systems have substantial performance parameters which should be considered. Firstly, ride Comfort is directly related to the acceleration sensed by vehicle passengers when traveling on a rough road. Secondly, Body motion, known as bounce, pitch, and roll of the sprung mass, is created primarily during cornering, breaking besides maneuvering. Then, Road handling is represented by the contact force of the tire and the road surface. Finally, suspension travel refers to the relative displacement between the sprung and the unsprung masses [1-4].

Generally, passive, semi-active and active suspension systems have been utilized to improve ride comfort of vehicles and their effectiveness have also been demonstrated. In case of passive suspension, it is not easy to improve dynamics behavior as the problem of passive system is if it designed heavily damped, road disturbance will be transferred to the vehicle body. On the other hand, if it considered softly damped, the vehicle will have a poor stability during cornering and maneuvering. Besides, the performance of the passive suspension depends on the road profile. In contrast, the active suspension can give better performance of suspension because it has the ability to adjust itself continuously to adapt with road condition. Better performance can be achieved by changing the actuator force in order to respond with road input variation.

Whereas a several number of researchers proposed theoretical and experimental studies describing the behavior of passive, semi-active and active suspension system as well. Abdel-Tawwab [5] presented a theoretical evaluation for twin accumulator suspension system and the results obtained were compared to those given by conventional passive suspension system. As a result, the twin accumulator system gave worthwhile improvements when compared to conventional passive suspension. In order to improve vehicle oscillation, Smith and Wang [6] proposed a comparative study of several simple passive suspension struts in which each containing at most one damper and inerter to improve both ride comfort and handling. They demonstrated that the implemented systems based on a quarter-car model had preferable attitude compared with a conventional passive suspension strut. Pable and Seshu [7] attempted to find the best parameters of a passive system which provides a performance as close to an active system as possible. Optimized passive settings were then obtained using least squares sense to generate equal suspension force as that of active case. Bo et al [8] developed a twin-accumulator hydro-pneumatic suspension based on the off-road vehicle and thus the working principles and elements construct of the developed suspension were studied. And then, they built a mathematical model of the developed suspension. The ride comfort of the vehicle with the developed suspension was analyzed theoretically compared with a single accumulator hydro-pneumatic suspension in both time domain and frequency domain.

Hong Biao and Nan [9] formulated an optimal vehicle suspension design problem with a quarter-car vehicle dynamic model. An optimization was conducted using Genetic algorithm technique. The proposed technique was a global optimization technique which was used to find an optimal design that minimizes an objective function subject to constraints. Mehmood et al [10] developed a mathematical model of quarter and a half car with a complete state space analysis and simulated by using MATLAB platform. The damping characteristic of the suspension damper was optimized in accordance with other vehicle dynamics parameter like sprung mass, unsprung mass, tire stiffness and damping, suspension stiffness and road input. Tran and Hasegawa [11] implemented a new design of passive suspension within component element named "inerter" to reduce sprung mass displacement and tire deflection in quarter-car model. The submitted design was optimized based on the minimization of cost functions for displacement, tire deflection with constraint function of suspension deflection limitation as well as the energy consumed by the inerter. Drehmer et al [12] developed a model of a vehicle with eight degrees of freedom including driver's seat under a random road profile. Here, particle swarm optimization and sequential quadratic programming algorithms were utilized to obtain the suspension optimal parameters at different conditions. They aimed to reduce the weighted RMS vertical acceleration based on human sensitivity curves. Recently, researchers give more attention in applying active suspension using several control methods in order to improve vehicle dynamic performance. Khajavi et al and Agharkakli et al developed an investigation to design an active suspension for a passenger car by designing LQR controller, which improves performance of the system with respect to design goals compared to passive suspension system [13, 14]. Choudhury and Sarkar [15] studied the performance of passive and active suspension system with PID controller. Therefore, Mathematical modeling of quarter car was done based on a two degree of freedom system. Rao [16] carried out a study to investigate the performance of a quarter car semi-active suspension system using PID controller under MATLAB Simulink Model considering the dynamic system used in this study was a linear system. In his work, two types of road profiles were used as input for the system. Results showed that the performance of body displacement and wheel displacement could be improved by using the proposed PID controller.

In this paper, a mathematical modeling of passive, twin-accumulator and active twin-accumulator with PI controller are established. The aim is to study the behavior of the previous systems based on different road condition as well as implement an optimization technique called evaluation strategy attached to the twin-accumulator system. Finally, PI controller is selected to be applied on the actuator force of the active system and thus the results are analyzed.

This paper is organized as follows; brief note on the twin-accumulator suspension is presented in section 2 while the mathematical modeling of a quarter car suspension model is built in section 3. The road profiles are explained in section 34. In section 5, Simulink structure of suspension models are discussed then the performance criteria, as well as the model parameters, are listed in section 6. Whereas, applied PI controller is in section 8 after that the results is presented in Section 9. Finally, the paper is concluded in section 10.

## 2. Twin-Accumulator Suspension Configuration:

Hydro-pneumatic suspension involves accumulator (flexible components) to generate spring force and a remote valve block (damping components) to generate damping force. A hydraulic cylinder replaces the damper strut and springs of the vehicle [17]. The twin-accumulator system consists of a parallel network of an accumulator and a throttle valve in series with another accumulator then this series network is parallel with second throttle valve as presented in Fig. 1.

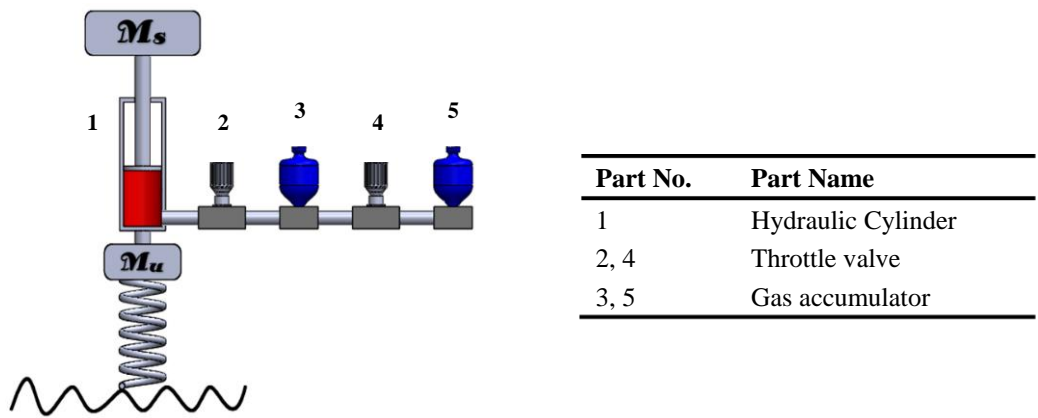

| Part No. | Part Name |
|---|---|
| 1 | Hydraulic Cylinder |
| 2, 4 | Throttle valve |
| 3, 5 | Gas accumulator |

Fig. 1. Twin-accumulator system construction.

## 3. Mathematical Modeling

In this work, a quarter car models with two degrees-of-freedom are carried out instead of full model based on laws of mechanics. It not only leads to simplify the analysis but also represents most of the features of the full model. Simulation software MATLAB/SIMULINK has been used to analyze the model.

### 3.1. Modeling assumptions:

Before the establishment of this model, three assumptions in the following should be required:

a) Vehicle near the equilibrium position to make a slight vibration, the spring-elastic element is a linear function of its displacement; the damping force is a linear function of its speed.

b) The operating conditions are driving at a constant speed on the smooth straight road.

c) Symmetrical to its longitudinal axis line of cars.

### 3.2. Conventional suspension system:

Fig. 2.a Shows a quarter car vehicle conventional hydro-pneumatic suspension system. Car body is denoted as sprung mass and the tire is denoted as un-sprung mass. Single wheel and axle is connected to the quarter portion of the car body through a spring and damper. The tire is assumed to have only the spring feature and is in contact with the road terrain at the other end. The road terrain serves as an external disturbance input to the system.

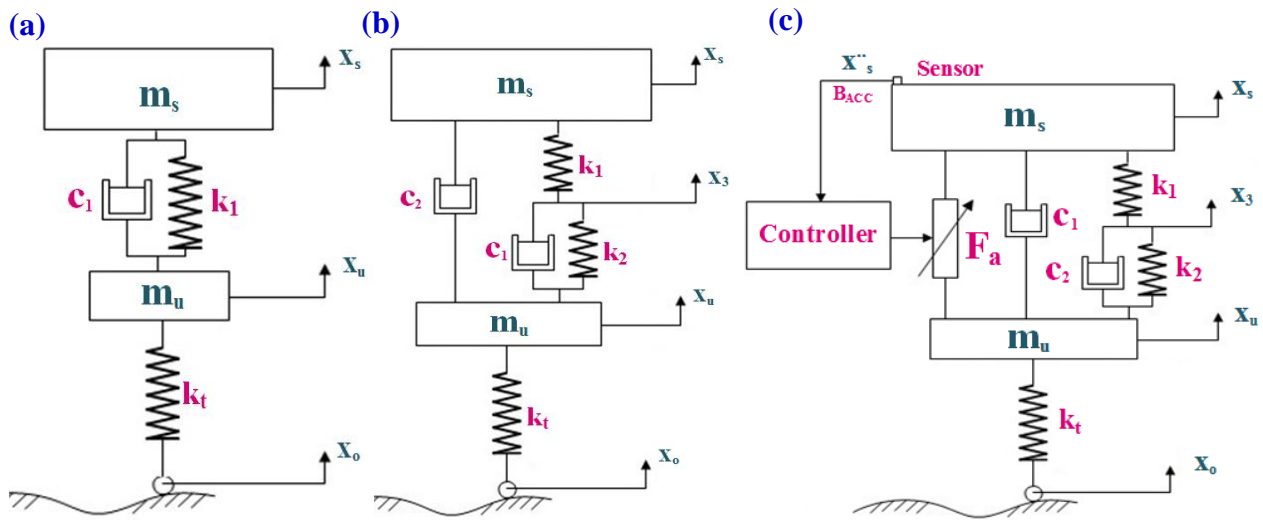

**Fig. 2. Quarter suspension model; (a) passive suspension; (b) twin accumulator suspension system; (c) active twin-accumulator model.**

Based on Newtonian mechanics the equations of the motion for the passive suspension system are given:

$$m_u \ddot{x}_u = k_t(x_o - x_u) - k_1(x_u - x_s) - c_1(\dot{x}_u - \dot{x}_s) \tag{1}$$

$$m_s \ddot{x}_s = k_1(x_u - x_s) + c_1(\dot{x}_u - \dot{x}_s) \tag{2}$$

### 3.3. Twin-Accumulator Suspension System:

The mathematical model of twin-accumulator suspension based on quarter vehicle model is presented in Fig. 2.b and the equation of motion can be written as:

$$m_s \ddot{x}_s = k_1(x_u - x_3) + c_1(\dot{x}_u - \dot{x}_s) \tag{3}$$

$$k_1(x_3 - x_s) = c_2(\dot{x}_u - \dot{x}_3) - k_2(x_u - x_3) \tag{4}$$

$$m_u \ddot{x}_u = k_t(x_o - x_u) - c_1(\dot{x}_u - \dot{x}_s) - c_2(\dot{x}_u - \dot{x}_3) - k_2(x_u - x_3) \tag{5}$$

### 3.4. Active twin-accumulator suspension system:

Fig. 2.c shows the equivalent theoretical model to an active suspension system with an actuator. Therefore, equation of motion that describe this system is established as:

$$m_s \ddot{x}_s = c_1(\dot{x}_u - \dot{x}_s) + k_2(x_u - x_3) + F_a \tag{6}$$

$$k_1(x_3 - x_s) = c_2(\dot{x}_u - \dot{x}_3) - k_2(x_u - x_3) \tag{7}$$

$$m_u \ddot{x}_u = k_t(x_o - x_u) - c_1(\dot{x}_u - \dot{x}_s) - c_2(\dot{x}_u - \dot{x}_3) - k_2(x_u - x_3) - F_a \tag{8}$$

### 4. Road Profile Excitation

In this paper, two kinds of road input are simulated. Firstly, Random road profile is used to excite the vehicle suspension system. A white noise road input signal should be accurately reflecting the real road condition when a vehicle drives on the road. Numerous researchers show that when the vehicle speed is constant, the road roughness is a stochastic process which is subjected to Gauss distribution, and it cannot be described accurately by mathematical relations. The vehicle speed power spectral density is a constant, which correspond with the definition and statistical characteristic of the white noise so that it can be simply simulated as road roughness time domain model. It is found that there are several ways to generate road elevation time domain model, such as filtering white noise generation method, random sequence generation method, filtering superposition method, AR (ARMA) method and fast Fourier inverse transform generation method (IFFT) [18]. Here, white noise generation method has been selected as it has a clear physical meaning and it is easy to be simulated.

Matching with the ISO/TC108/SC2N67 international standard, it is recommended that the power spectral density of road vertical elevation (PSD), $G_q$ ($n_o$), described as the following formula as a fitting expression:

$$G_q(n) = G_q(n_o)(\frac{n}{n_o})^{-w} \qquad (9)$$

Taking into consideration the random behavior of the white noise method, the transformation of white noise signal can perfectly simulate the actual pavement condition. Additionally, it is used for the vehicle excitation input of road roughness, which is always use the road power spectral density to describe its statistical properties.

International Organization for Standardization provides classifications of road roughness using Power Spectral Density values as in Table 2. Class D road profile was selected is this work to be the main road profile. The road input was described in reference [18] as:

$$\frac{q(s)}{w(s)} = \frac{2\pi n_o \sqrt{G_q(n_o)v}}{s + 2\pi f_o} \qquad (10)$$

Where w(t) is Gaussian white noise filter. In this study, V= 20 m/s, $G_q$ ($n_o$)= 1024e-6 m$^3$, $n_o$= 0.1 m$^{-1}$ and $f_o$= 0.

Table 2. Eight degrees of road profile roughness.

| Road Level | $G_q(n_o)/(10^{-6} m^3)$ $(n_o = 0.1\ m^{-1})$ | $\sigma_q(n_o)/(10^{-3} m)$ $0.011\ m^{-1} < n > 2.83\ m^{-1}$ |
|---|---|---|
| | Geometric Average | Geometric Average |
| A | 16 | 3.81 |
| B | 64 | 7.61 |
| C | 256 | 15.23 |
| **D** | **1024** | **30.45** |
| E | 4096 | 60.90 |
| F | 16384 | 121.80 |
| G | 65536 | 243.61 |
| H | 262144 | 487.22 |

The Simulink model of road disturbance is developed according to the mathematical equation (10) and the entire system is simulated in the Simulink as shown in Fig. 3. Then the road profile attitude is presented in Fig. 4.a and Fig. 4.b.

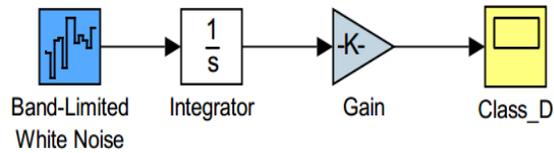

**Fig. 3. White noise road Simulink model.**

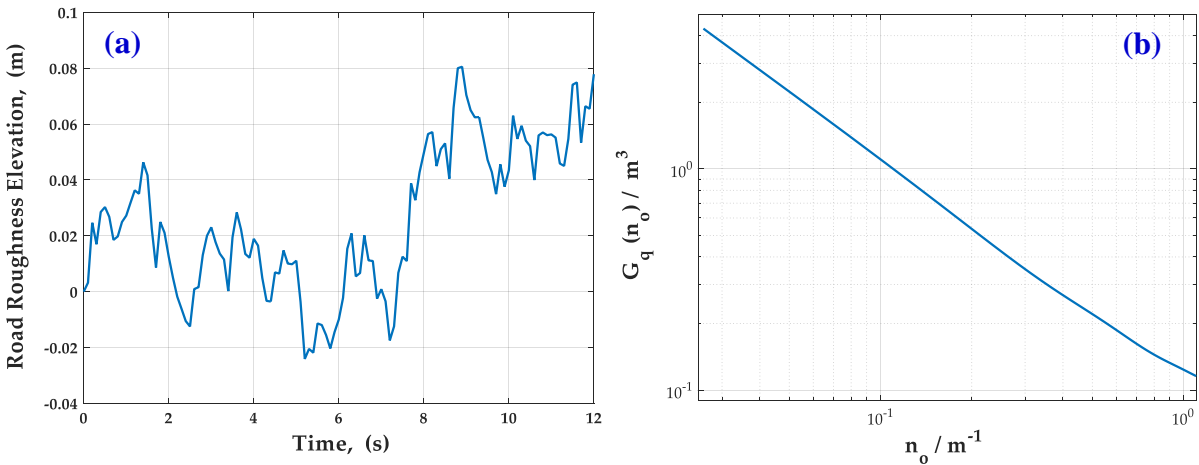

**Fig. 4. (a) Road input disturbance for D-Class; (b) Log. plot in frequency domain for D-class road.**

Secondly, as can be seen in Fig. 5, the systems are also excited by a road disturbance of step input with the height of 0.02 m at t=2 sec.

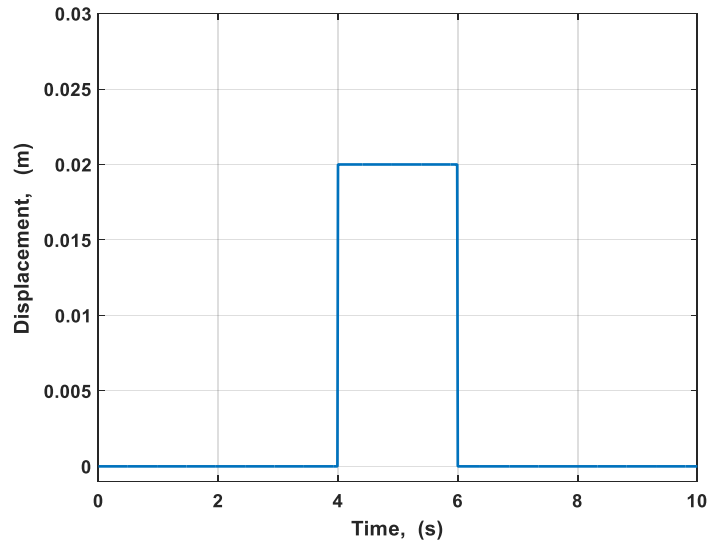

**Fig. 5. Road bump displacement.**

## 5. Simulink implementation of Suspension models

Quarter suspension models for passive, twin and PI controlled system are implemented and simulated using MATLAB/SIMULINK. Fig. 6 present the interface of the tested systems. The models were excited by a random input of Class-D and step input and thus results have been recorded and discussed.

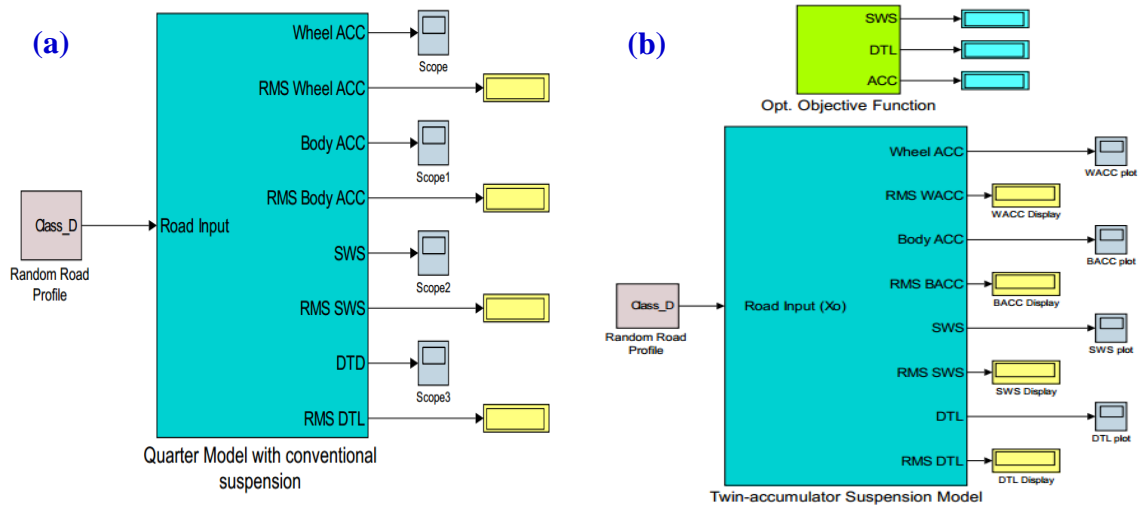

**Fig. 6. Simulink of conventional suspension system; (b) Simulink of twin-accumulator system with optimization model based on multi-objective function.**

## 6. Performance criteria and model parameters

The performance of each suspension system can be assessed quantitatively in terms of three parameters. These are chosen to represent each of the conflicting requirements of the suspension and have been widely used and accepted as a measure of system performance [1, 19]. The main model parameters are listed in Table 3.

### 3.1. Discomfort (ACC)

This provides a single number index of system performance in terms of ride quality. It can be defined as the R.M.S value of frequency weighted vertical body acceleration.

### 3.2. Suspension Working Space (SWS)

This parameter is defined as the RMS value of wheel to body displacement $x_u - x_s$ and measures the variation of the displacement about its static position.

### 3.3. Tire Loading Parameter (DTL)

This parameter is defined as the RMS value of tire load variations from the static value. DTL can be considered as a measure of road holding ability since a variation in the tire load results in a varying contact length and consequently a net reduction inside or braking force.

**Table 3. Quarter vehicle parameters**

| Parameters | Symbol | Value | Unit |
|---|---|---|---|
| Sprung mass | $m_s$ | 300 | kg |
| Unsprung mass | $m_u$ | 40 | kg |
| Frist Damping coefficient | $C_1$ | 1000 | N.s/m |
| Frist spring stiffness | $k_1$ | 15000 | N/m |
| Second Damping coefficient | $C_2$ | 1000 | N.s/m |
| Second spring stiffness | $k_2$ | 15000 | N/m |
| Tire stiffness | $k_t$ | 20000 | N/m |

## 7. Implementation of Optimization via Multi-objective function

In this paper, the aim of designing a multi-objective function in active suspension system is to minimize the difference between the body and tire displacements to provide more comfort against the road disturbance. The objective function has been considered as a combination of ride comfort and ride safety by taking appropriate weighting factors which are listed in Table 4.

Evaluation strategy technique is carried out to find an optimal solution expressed by the objective function under the fulfillment of constraint conditions. Fig. 7. illustrates the developed objective function Simulink model. Table 5. states the lower and upper boundaries of designed parameters.

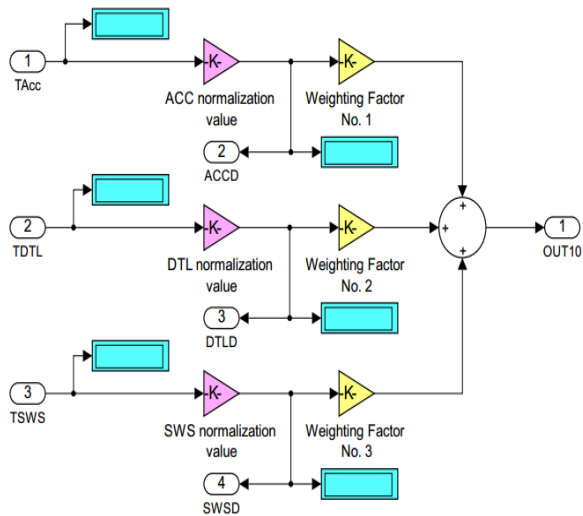

**Table 4. considered weighting factors.**

| Variable | BACC | DTL | SWS |
|---|---|---|---|
| Weighting Factor | 0.5 | 0.45 | 0.05 |

Fig. 7. Multi-objective function implementation

**Table 5. upper and lower boundary of parameters.**

| Parameter | Upper boundary | Lower boundary |
|---|---|---|
| $C_1$ | 2500 | 900 |
| $k_1$ | 40000 | 15000 |
| $C_2$ | 2500 | 900 |
| $k_2$ | 40000 | 15000 |

The selected optimization technique (Evaluation strategy technique) is applied and thus the simulation is carried out with 5, 10 and 50 iterations then the results are presented and discussed. Considering that, the initial and optimum values of the mentioned parameters over the two-road excitation is illustrated in Table 6 and Table 7.

| Table 6. Optimal system parameters for random input | | | | Table 7. Optimal system parameters for step input | | | |
|---|---|---|---|---|---|---|---|
| Parameter | Initial Value | Optimal Value | Unit | Parameter | Initial Value | Optimal Value | Unit |
| $c_1$ | 1000 | 900 | N.s/m | $c_1$ | 1000 | 900 | N.s/m |
| $k_1$ | 15000 | 15000 | N/m | $k_1$ | 15000 | 15000 | N/m |
| $c_2$ | 1000 | 900 | N.s/m | $c_2$ | 1000 | 2196 | N.s/m |
| $k_2$ | 15000 | 18801 | N/m | $k_2$ | 15000 | 40000 | N/m |

## 8. PI controller

The benefit of controlled suspension is that a better set of design trade-offs are possible to be compared with passive suspension. The mathematical representation of simplified PI controller scheme is given by:

$$G_c = K_p e(t) + K_i \int e(t)dt \tag{11}$$

Where $G_c$ is the controller output, $k_P$ is proportional gain, $k_i$ is integral gain, e(t) is input to the controller and e(t) dt is the time integral of the input signal. Fig. 8.a presents the closed loop diagram of the selected controller. For evaluating the performance of the applied controller, the simulations are carried out on the active twin-accumulator suspension model.

In this research, a unity feedback circuit is formed depending on the body acceleration as the output variable. Taking the set point as zero, the error signal is fed to the controller. The closed loop block diagram is shown in Fig. 8.b.

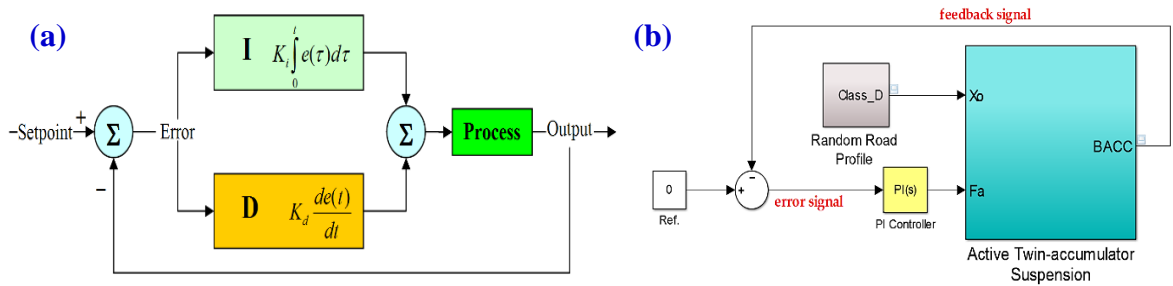

**Fig. 8. (a) Block diagram of a PI controller in a feedback loop; (b) Simulink of active suspension system with PI controller.**

Firstly, the proportional and integral terms are summed with initial values to calculate the output of the PI controller. Then, auto tuning for PI controller has been applied using robust response tuning method with the MATLAB/Simulink software. Both initial and tuned values of the controller gains are shown in Table 8. The introduced tuning method is a trial and error method with a simple and common step due to determine the best value.

**Table 8. PI controller gains**

| Controller Gains | Initial Value | Tuned Value Random Input | Tuned Value Step Input |
|---|---|---|---|
| $k_p$ | 0 | 0 | 0 |
| $k_i$ | 1 | 1904 | 800 |

## 9. Simulation and Results

The simulation tests are conducted for three suspension modes: passive suspension, conventional optimized twin-accumulator suspension as well as active twin-accumulator suspension with PI controller. The first case of road disturbance is a random profile based on white noise method.

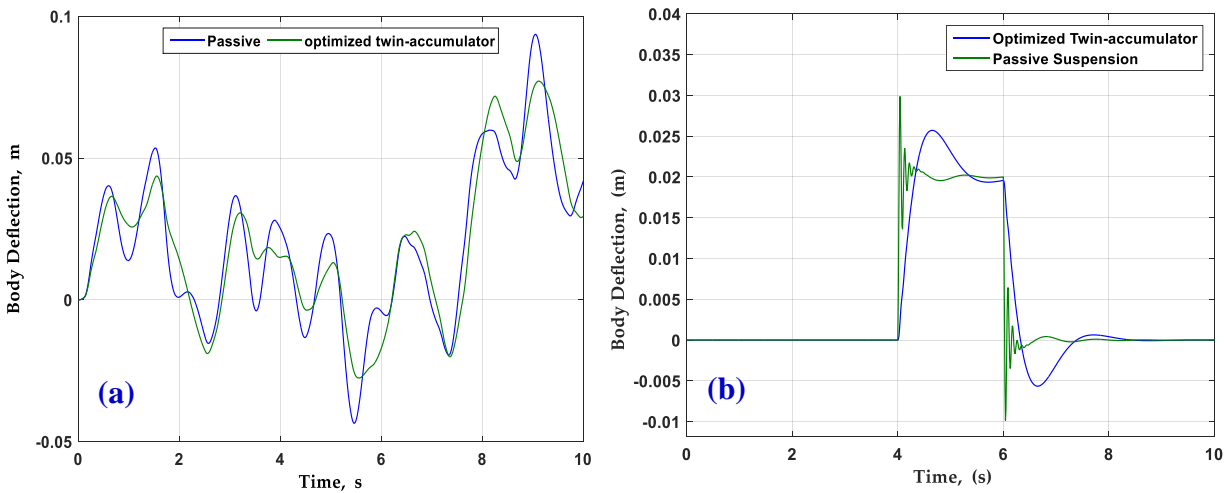

**Fig. 9. (a) Time response of vehicle body defection over random input; (b) Time response of vehicle body defection over random input.**

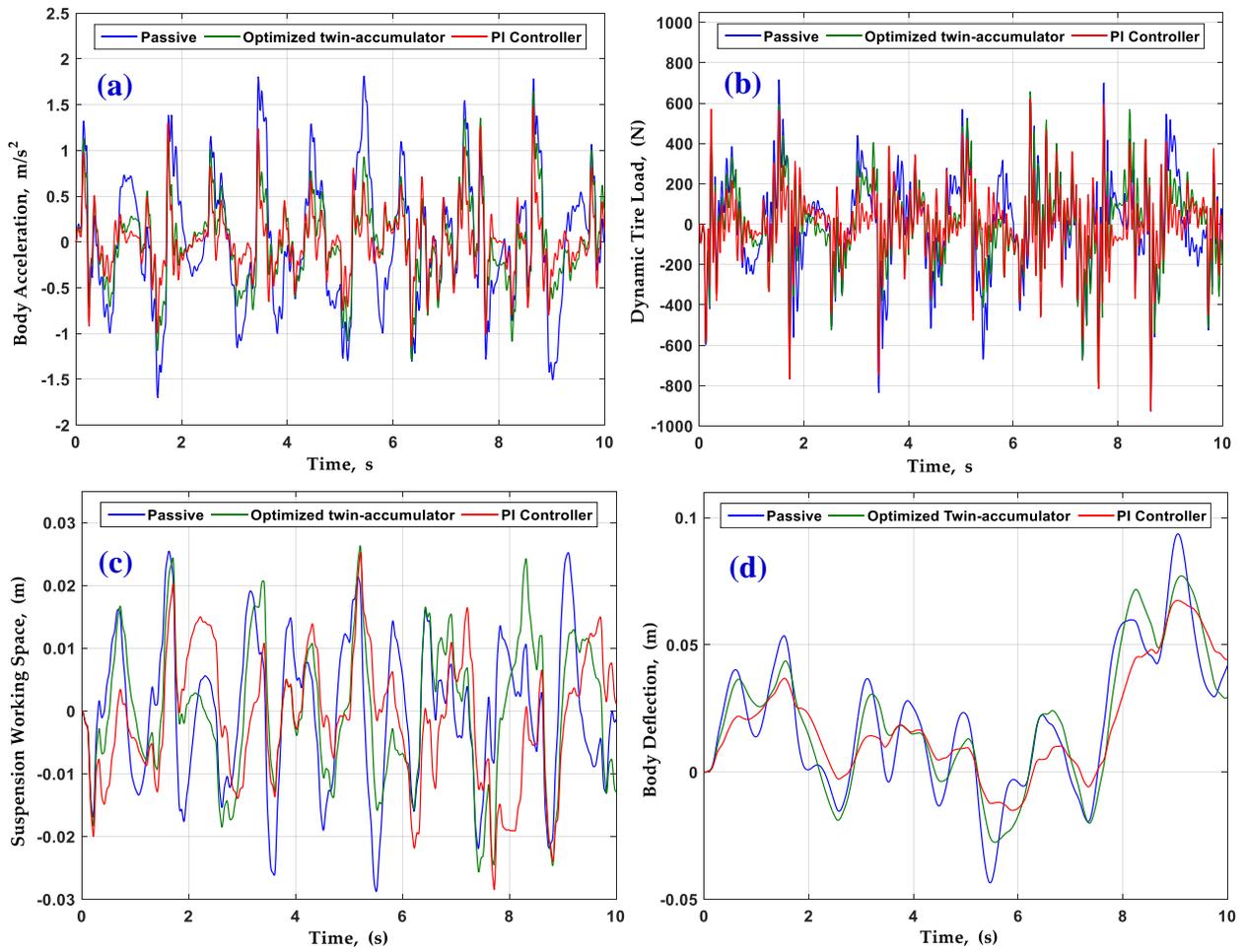

**Fig. 10. Suspension performance (a) Body acceleration response; (b) Dynamic tire load response;**

**(c) Suspension travel response; (d) Body displacement response.**

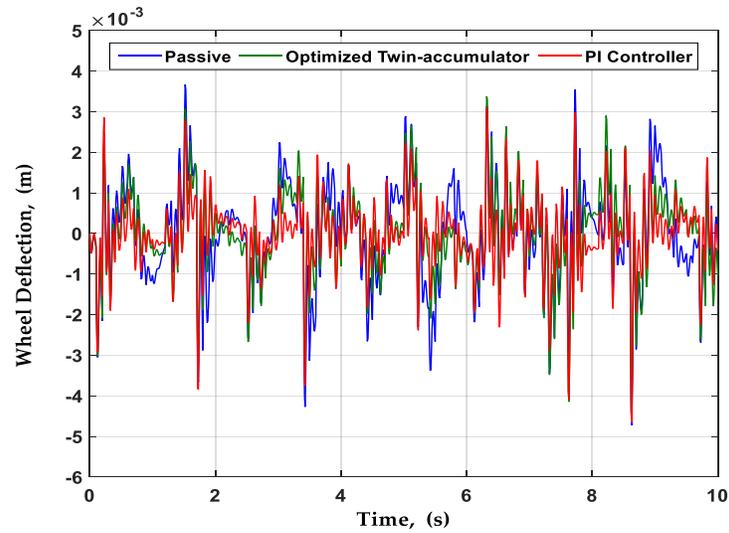

**Fig. 11. Wheel deflection response of suspension models.**

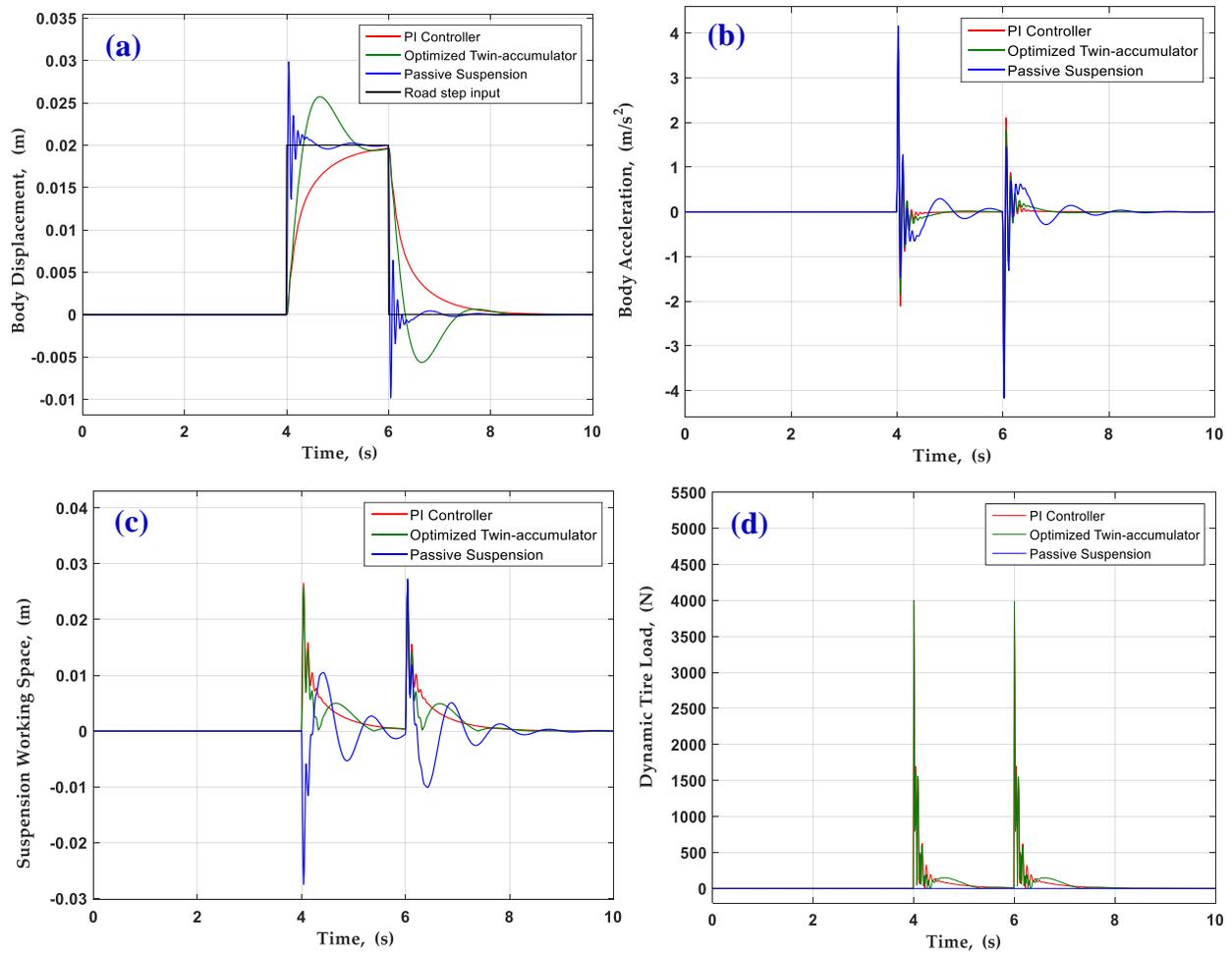

**Fig. 12. Suspension performance for step road input. (a) Body displacement response; (b) Body acceleration response; (c) Suspension travel response; (d) Dynamic tire load response.**

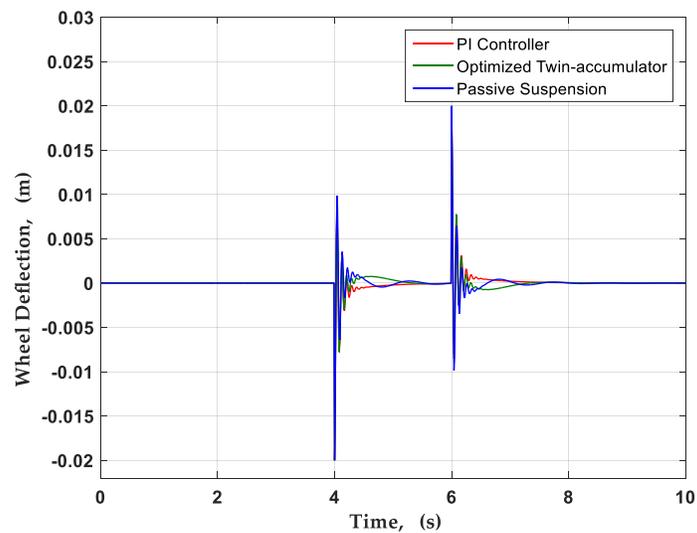

**Fig. 13. Wheel deflection response of step road input.**

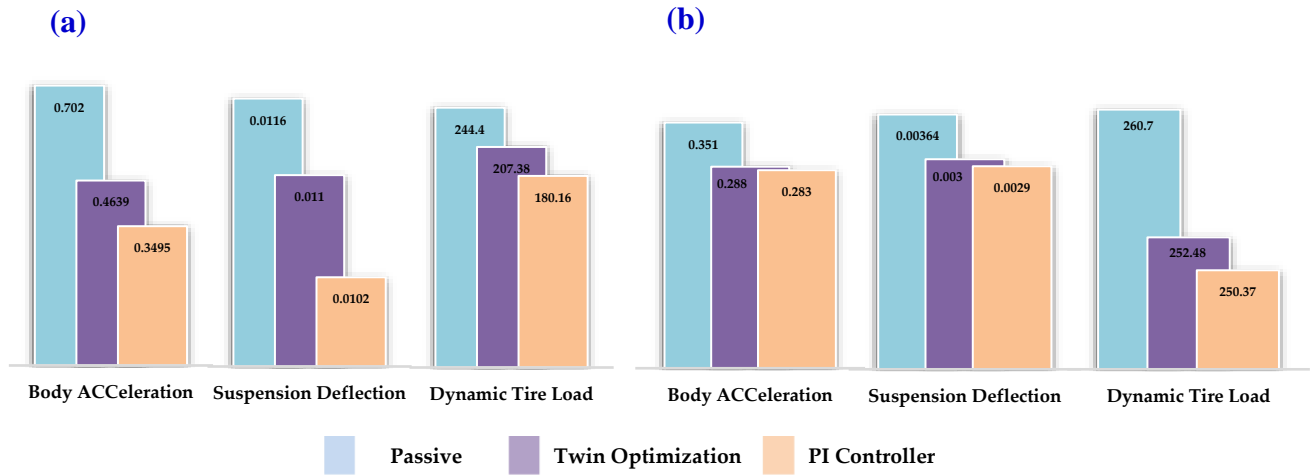

Fig. 14. (a) RMS values of performance index over random road; (b) RMS values for step input.

Fig. 9 show that the body deflection is less for the optimized twin-accumulator suspension than the deflection for the passive suspension in both random and step road profile. After which the controller is applied, the system performance achieved more improvement as indicated in Fig. 10 and 12. Another behavior category, Fig. 11 and 13 indicate the wheel displacement response for the two road inputs. It can be seen that the results for optimized twin-accumulator are closely matched with to PI controlled system. Consequently, the dynamic tire load which is shown in Fig. 10.b and 12.d have the same attitude as the wheel deflection. While the passive suspension system is worse than the other systems.

As can be observed in Figs. 10.c and 12.c, the suspension working space of the PI controller system is lower than the twin-accumulator and passive as well. In Figs. 10.a and 12.b, the body acceleration of the active system with PI controller is more beneficial than the other systems.

## 10. Conclusion

All in all, in this work, an optimization technique is carried out on the twin accumulator suspension system so the multi-objective function is developed to minimize the performance criteria of the system. The PI controller has been successfully implemented in an active twin-accumulator suspension system through simulation analysis. After that, a comparison between passive, optimized twin-accumulator and PI controller is presented as well as their performances are analyzed. The compassion depends on the RMS of the body acceleration, dynamic tire load, and suspension travel. The performance of twin-accumulator active suspension system with PI controller has been proven to perform better than the passive suspension system.

## 11. Proposal for future work

The recommendation for future work is to expand the vehicle dynamic model to a half model with an active twin-accumulator suspension system to study the pitch attitude. After that, an advanced control strategy should be designed and expanded to include the body acceleration, the suspension travel, and the wheel deflection. Implement a regenerative energy harvesting suspension system depending on the hydraulic based system [20].

**Notations**

| Symbol | Definition | Unit |
| --- | --- | --- |
| $m_s$ | Sprung mass | kg |
| $m_u$ | Unsprung mass | kg |
| $C_1$ | Frist Damping coefficient | N.s/m |
| $C_2$ | Second Damping coefficient | N.s/m |
| $k_1$ | Frist spring stiffness | N/m |
| $k_2$ | Second spring stiffness | N/m |
| $k_t$ | Tire stiffness | N/m |
| $x_s$ | Sprung mass displacement | m |
| $\ddot{x}_s$ | Body acceleration | m/s² |
| $x_u$ | Unsprung mass displacement | m |
| $\ddot{x}_u$ | Wheel acceleration | m/s² |
| $F_a$ | Actuator force | N |
| $V$ | Vehicle speed | m/s |
| $q(t)$ | Road profile excitation | |
| $w(t)$ | Gaussian white noise filter | |
| $k_p$ | Proportional controller gain | |
| $k_i$ | Integral controller gain | |

# Appendix

Final interface of the developed Simulink model

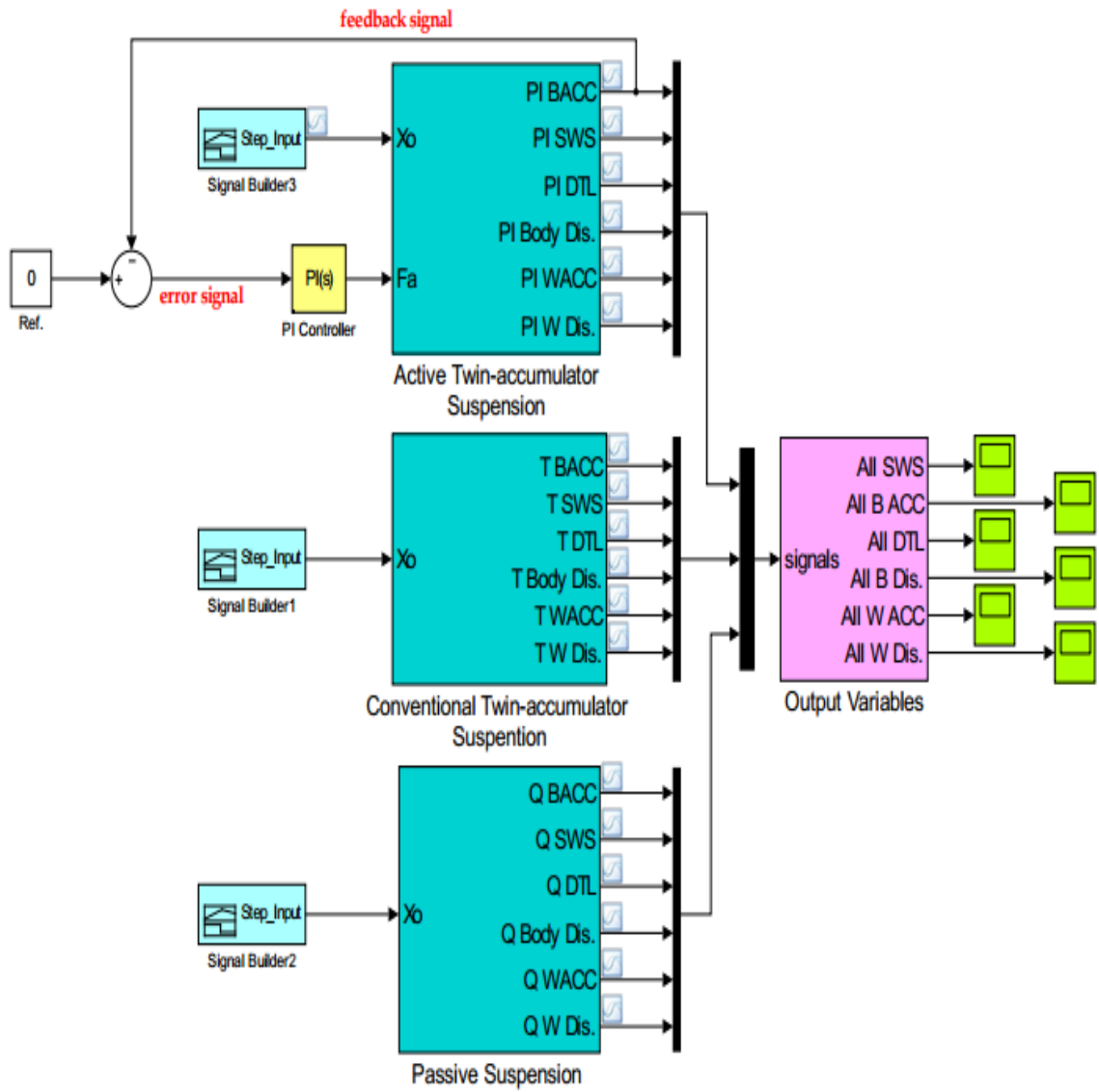